\documentclass[preprint,aps,pre,amsmath,amssymb,amsfonts,showpacs,showkeys,floatfix]{revtex4-1}

\usepackage[dvips]{graphicx}
\usepackage{hyperref}
\usepackage{color}

\newcommand{\articletype}{paper}

\newcommand{\rmd}{\mathrm{d}}

\newcommand{\naturalset}{\mathbb{N}}

\newcommand{\ie}{{i.e.}}
\newcommand{\eg}{{e.g.}}

\newcommand{\etal}{\textit{et al.}}

\newcommand{\Heaviside}[1]{\Theta( #1 )}

\newcommand{\diff}[3]{\dfrac{\rmd^{ #1 } { #2 }}{\rmd { #3 }^{ #1 }}}

\newcommand{\average}[1]{\langle #1 \rangle}

\newcommand{\withindex}[3]{{{ #1 }^{ #2 }_{ #3 }}}
\newcommand{\vecwithindex}[3]{\withindex{\boldsymbol{ #1 }}{ #2 }{ #3 }}

\newcommand{\systime}[2]{\withindex{t}{ #1 }{ #2 }}
\newcommand{\cylr}[2]{\withindex{r}{ #1 }{ #2 }}
\newcommand{\cylangle}[2]{\withindex{\theta}{ #1 }{ #2 }}
\newcommand{\cylz}[2]{\withindex{z}{ #1 }{ #2 }}

\newcommand{\mom}[2]{\withindex{p}{ #1 }{ #2 }}

\newcommand{\normalvec}[2]{\vecwithindex{n}{ #1 }{ #2 }}
\newcommand{\posvec}[2]{\vecwithindex{r}{ #1 }{ #2 }}
\newcommand{\velvec}[2]{\vecwithindex{v}{ #1 }{ #2 }}

\newcommand{\forcevec}[2]{\vecwithindex{f}{ #1 }{ #2 }}

\newcommand{\overlap}[2]{\withindex{\xi}{ #1 }{ #2 }}
\newcommand{\mass}[2]{\withindex{m}{ #1 }{ #2 }}
\newcommand{\diameter}[2]{\withindex{d}{ #1 }{ #2 }}
\newcommand{\viscosity}[2]{\withindex{\eta}{ #1 }{ #2 }}
\newcommand{\springcoeff}[2]{\withindex{k}{ #1 }{ #2 }}
\newcommand{\numps}{{N}}
\newcommand{\numwps}{{N_\mathrm{w}}}
\newcommand{\setps}{{\Pi_\mathrm{p}}}
\newcommand{\setwps}{{\Pi_\mathrm{w}}}
\newcommand{\settotalps}{{\Pi}}
\newcommand{\elforcevec}[1]{\forcevec{\mathrm{el}}{ #1 }}
\newcommand{\visforcevec}[1]{\forcevec{\mathrm{vis}}{ #1 }}
\newcommand{\wavelength}{{\lambda}}
\newcommand{\avgtuberadius}{{a}}
\newcommand{\perisamplitude}{{b}}
\newcommand{\perisvel}{{c}}
\newcommand{\minwidth}{{w}}
\newcommand{\upperminwidth}{{w_\mathrm{max}}}
\newcommand{\critminwidth}{{w_\mathrm{c}}}
\newcommand{\newcritminwidth}{{w_\mathrm{c}^\prime}}
\newcommand{\strainrate}{{\dot{\epsilon}}}
\newcommand{\massflux}{{J}}

\newcommand{\maxmassflux}{{\tilde{J}}}
\newcommand{\stmassflux}{{J_\mathrm{st}}}
\newcommand{\fluidstmassflux}{{J_\mathrm{st}^\mathrm{fluid}}}
\newcommand{\tubelength}{{L}}
\newcommand{\transtime}{{\tau}}
\newcommand{\transtimeexp}{{\alpha}}
\newcommand{\transtimefluc}{{\chi_\tau}}

\newcommand{\dens}{{\rho}}
\newcommand{\effdens}{\bar{\rho}}
\newcommand{\dynsuscept}[2]{{\chi^{ #1 }_{ #2 }}}
\newcommand{\maxdynsuscept}[1]{{\dynsuscept{\mathrm{max}}{ #1 }}}

\begin{document}

\title{Phase transition in peristaltic transport of frictionless granular particles}
\date{\today}

\author{Naoki Yoshioka}
\email{naoki@yukawa.kyoto-u.ac.jp}
\affiliation{Yukawa Institute for Theoretical Physics, Kyoto University, Kitashirakawa Oiwake-cho, 606-8502 Kyoto, Japan}

\author{Hisao Hayakawa}
\affiliation{Yukawa Institute for Theoretical Physics, Kyoto University, Kitashirakawa Oiwake-cho, 606-8502 Kyoto, Japan}



\begin{abstract}
Flows of dissipative particles driven by the peristaltic motion of a tube
  are numerically studied.
A transition from a slow ``unjammed'' flow to a fast ``jammed'' flow is found
  through the observation of the flow rate at a critical width of the bottleneck of a peristaltic tube.
It is also found that the average and fluctuation of the transition time,
  and the peak value of the second moment of the flow rate
  exhibit power-law divergence near the critical point and
  that these variables satisfy scaling relationships near the critical point.
The dependence of the critical width and exponents on the peristaltic speed
  and the density is also discussed.
\end{abstract}

\maketitle

\section{Introduction}\label{sec:introduction}

Peristalsis is a progressive wave of area contraction and expansion in a tube.
Transport due to the peristaltic motion of  a tube is
  one of the main transport mechanisms in biological systems
  such as the esophagus, small intestine, ureter, and so forth \cite{jaffrin:1971}.
Peristaltic transport is also found in the pumping of fluids, 
  known as peristaltic pumps,
  which are employed in medical and food engineering fields \cite{jaffrin:1971}.

The study of peristaltic transport has a long history
  and was particularly motivated by the desire to  understand the transport of fluids in the ureter.
Peristaltic transport in a Stokes fluid
  was studied in the late 1960s \cite{burns:1967, shapiro:1969}
  and early 1970s \cite{li:1970, yin:1971, weinberg:1971}.
A study on
   the peristaltic transport of a micropolar fluid \cite{srinivasacharya:2003}
  following this direction has also been reported.
In contrast, there have been few studies
  on the transport of particles under a peristaltic condition
  such as single particles \cite{hung:1976, fauci:1992}
  and dilute passive solid particles
  \cite{jimenez-lozano:2009, jimenez-lozano:2010}
  suspended in a Newtonian fluid.
Moreover, to the best of our knowledge, the peristaltic transport of dense particles has never been studied,
  even though such a situation is commonly observed in flows of red blood cells,
  in the peristaltic pumping of corrosive sand and solid foods, and so forth \cite{jaffrin:1971}.

Here, we consider the peristaltic transport of dry and smooth granular particles,
  which is closely related to the transport of particles through a bottleneck
  \cite{helbing:2001, nakajima:2009},
  particularly the discharge of grains from a silo
  \cite{le-pennec:1996, longhi:2002,
        beverloo:1961, nedderman:1982, mankoc:2007, de-song:2003, aguirre:2010,
        hou:2003, zhong:2006, huang:2006, janda:2009,
        to:2001, to:2005, zuriguel:2003, zuriguel:2005, janda:2008}.
  Through previous studies it has been established that
  there are three regimes depending on the linear size of the bottleneck, $w$.
If $w$ is sufficiently large, particles flow continuously
  \cite{beverloo:1961, nedderman:1982, mankoc:2007, de-song:2003, aguirre:2010}.
It is also known that
  the mass flow rate $Q$ satisfies the Beverloo law $Q \propto (w - w_0)^B$
  for an empirical value $w_0$,
  where $B$ is $3/2$ and $5/2$ for two- and three-dimensional systems
  under gravity, respectively \cite{beverloo:1961, nedderman:1982, mankoc:2007},
  and $B$ is $1$ for disks on a conveyor belt \cite{de-song:2003, aguirre:2010}.
With decreasing size of the bottleneck, 
  the flow becomes intermittent owing to the formation and breakdown of an arch
  at the outlet \cite{hou:2003, zhong:2006, huang:2006, janda:2009},
  and finally the flow stops, \ie, \textit{jamming} occurs
  \cite{to:2001, to:2005, zuriguel:2003, zuriguel:2005, janda:2008}.
Note, however, that the term jamming used in this context is
  slightly different from that in the \textit{jamming transition}
  of granular matter discussed in recent studies
  \cite{ohern:2002, ohern:2003, dauchot:2005, olsson:2007,
        hatano:2008, otsuki:2009a, otsuki:2009b, otsuki:2010}.
Indeed, the jamming transition can be observed
  even for smooth, \ie, frictionless, granular particles,
  while the jamming of a granular flow at the bottleneck is
  due to the persistent arch formation by grains,
  which can be observed only for rough, or frictional, particles.
What actually corresponds to the jamming transition in the discharge of grains
  appears to be the so-called dilute-to-dense transition
  \cite{hou:2003, zhong:2006, huang:2006}, which is observed at a phase boundary
  between continuous and intermittent regimes.

In this \articletype, we report the results of our simulations
  on the flow rate of a peristaltic granular flow.
We consider dry, smooth, monodisperse, and spherical particles 
  in a three-dimensional sinusoidally oscillating tube.
After the  introduction of our numerical model in Sec.\ \ref{sec:model},
  we present the results of our simulations in Sec.\ \ref{sec:results}.
Through the observation of the flow rate, we find a transition from a slow ``unjammed" flow to a fast ``jammed" flow occurs
  if the minimum width of the peristaltic tube is smaller than a critical value.
We also find that the average and fluctuation of the transition time,
  and the peak value of the second moment of the flow rate
  exhibit power-law divergence with nontrivial exponents
  near the minimum width.
Moreover, we verify the existence of scaling functions for these quantities.
It is also found that the critical width is almost independent of the density
  but depends on peristaltic velocity.
In Sec.\ \ref{sec:summary}, we discuss and summarize our results.

\section{Model}\label{sec:model}

\begin{figure}
  \begin{center}
    \includegraphics[height=.6\linewidth]{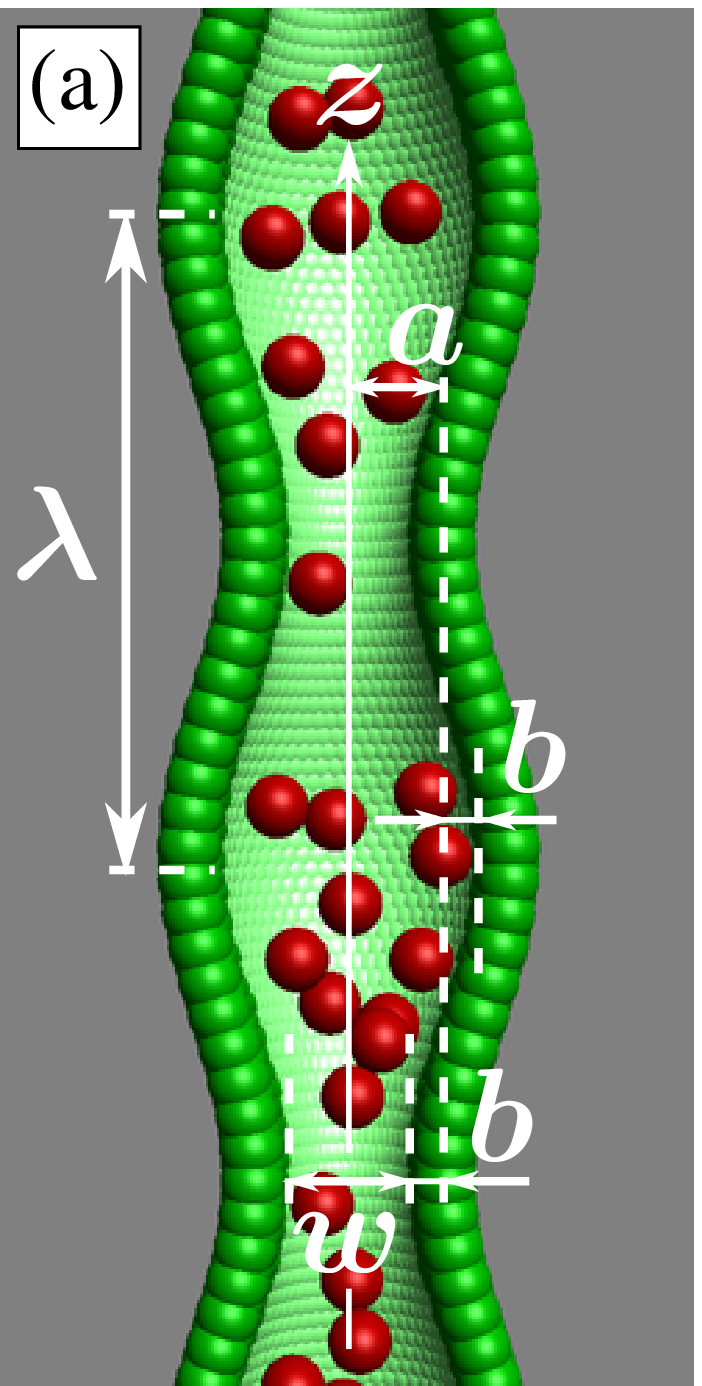}
    \includegraphics[height=.6\linewidth]{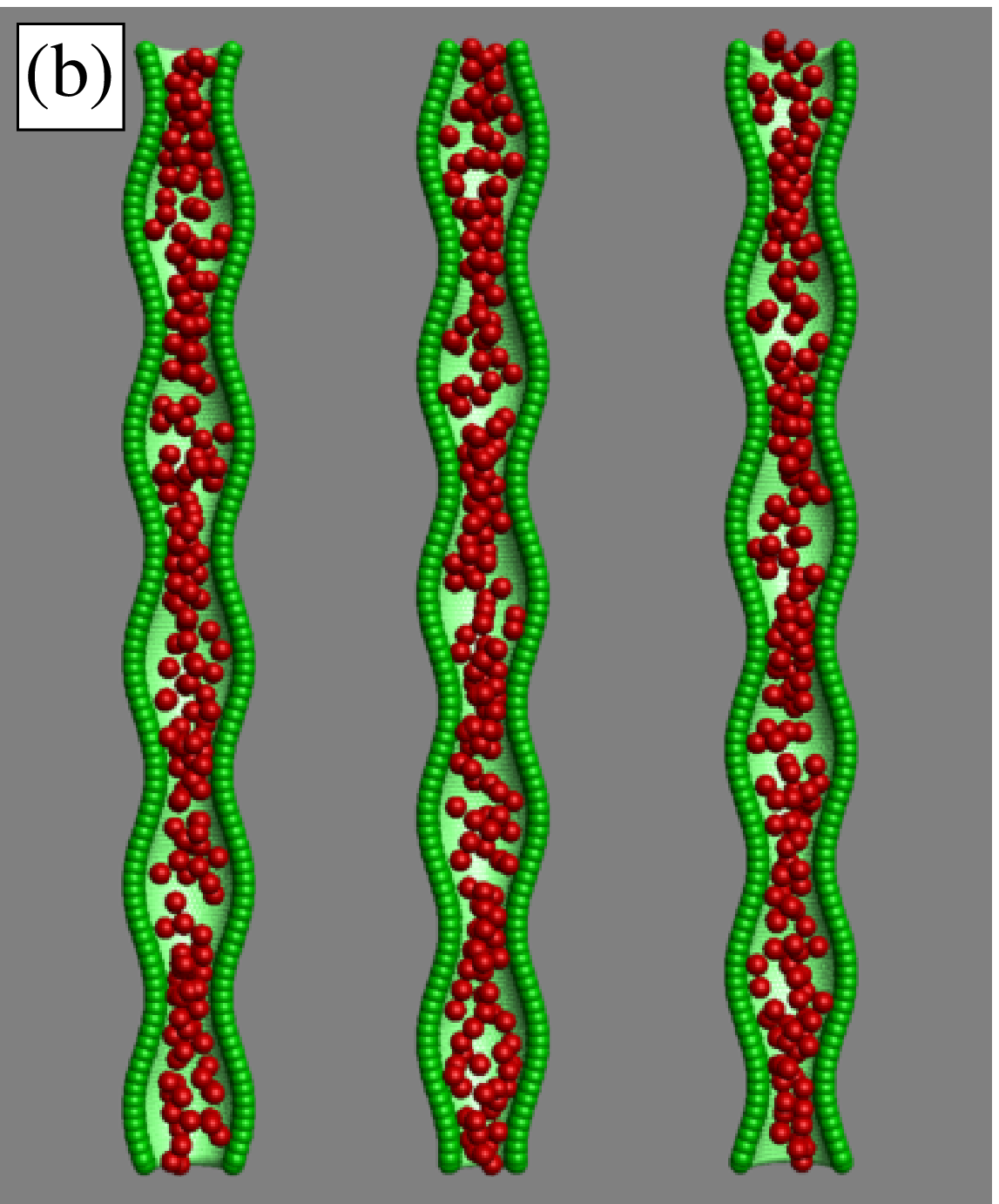}
    \includegraphics[height=.6\linewidth]{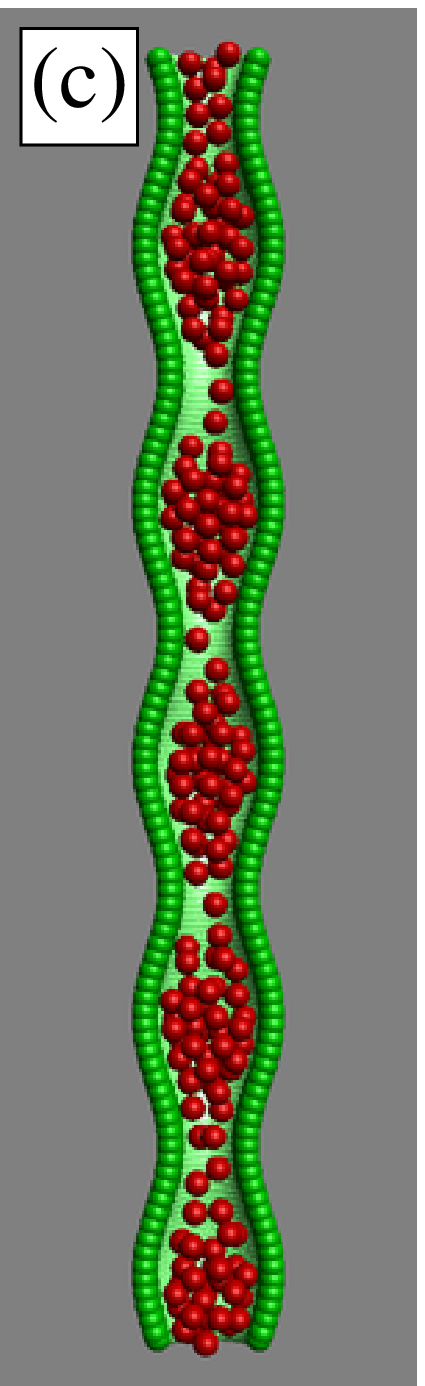}
  \end{center}
  \caption{\label{fig:model}
  (Color online)
  (a) Snapshot of our simulation showing some of the parameters in our model.
  See the text for details.
  (b) Sequential snapshots of the simulation
    for $\minwidth = 2.0$, $\strainrate = 1.83 \times 10^{-1}$,
    and $\effdens = 2.96 \times 10^{-1}$.
  Time increases from left to right,
    and the peristaltic wave of the wall progresses upward.
  The system remains in the unjammed flow regime when $\systime{}{} < \transtime$.
  (c) Snapshot of stationary state for the parameters used in (b)
    showing the jammed flow regime when $\systime{}{} > \transtime$.
  Dark (red) spheres and light (green) spheres are transported particles
    and particles embedded in the wall, respectively.
  Wall particles facing the reader are not visualized
    to show the transported particles clearly.
}
\end{figure}

  We adopt the three-dimensional molecular dynamics simulation for soft-core particles known as
   the discrete element method (DEM) 
  to simulate the peristaltic flow of spherical granular particles.
We assume that the particles are dry, smooth, and monodisperse spheres.
This means that we can ignore tangential forces that lead to the sliding and rotation of particles as well as adhesive forces.
 We also assume that all the material properties of the particles are identical.
Moreover, to idealize our setup, we ignore gravity and the hydrodynamic force through interstitial fluids such as air.
Peristaltic motion is represented by the propagation of a spatially oscillating wall along the axial direction
of a tube,
 which is described in detail in the following.

The system involves  $\numps$ monodisperse, smooth, and mobile granular spheres with
  diameter $\diameter{}{}$ and mass $\mass{}{}$
  in a peristaltic tube with length $\tubelength$
  under a periodic boundary condition along the direction of propagation $\cylz{}{}$ of peristaltic waves.
The wall of the tube consists of $\numwps$ embedded overlapping spheres.
We ignore interactions among the particles embedded in the wall and the deformation of the wall.
Therefore, we assume that the peristaltic wave of the wall is uniform and perfectly controllable.
See the schematic diagram shown in Fig.\ \ref{fig:model} (a).

We label a set of particles $\setps = \{ 1, \ldots, \numps \}$ denoting mobile particles and 
  $\setwps = \{ \numps + 1, \ldots, \numps + \numwps \}$ denoting particles embedded in the wall,
  where $\numwps$ is the number of particles embedded in the wall, with the relation $\settotalps = \setps \cup \setwps$.
For $i \in \Pi_{\mathrm{p}}$, \ie, mobile particles, the equation of motion is given by
\begin{equation}
  \mass{}{} \diff{2}{\posvec{}{i}}{\systime{}{}}
  = \sum_{j \in \settotalps, j\ne i}
    \bigl( \elforcevec{ij} + \visforcevec{ij} \bigr),
\label{eq:equation of motion}
\end{equation}
where $\elforcevec{ij}$ and $\visforcevec{ij}$ are respectively the elastic and viscous contact forces
acting on particle $i$ by particle $j$.
For their explicit forms, we adopt the following model:
\begin{align}
  \elforcevec{ij}
&
  = \springcoeff{}{} \overlap{}{ij}
    \Heaviside{\overlap{}{ij}} \normalvec{}{ij},
\\
  \visforcevec{ij}
&
  = -\viscosity{}{} (\velvec{}{ij} \cdot \normalvec{}{ij})
    \Heaviside{\overlap{}{ij}} \normalvec{}{ij},
\end{align}
  where $\Heaviside{x}$ is a step function
  with $\Heaviside{x} = 1$ for $x > 0$ and $\Heaviside{x} = 0$ otherwise,
  $\springcoeff{}{}$ denotes the spring constant,
  $\viscosity{}{}$ is the viscosity,
  $\overlap{}{ij}  = \diameter{}{} - |\posvec{}{ij}|$,
  $\posvec{}{ij} = \posvec{}{i} - \posvec{}{j}$,
  $\normalvec{}{ij}   = \posvec{}{ij} / |\posvec{}{ij}|$,
  and $\velvec{}{ij} = \velvec{}{i} - \velvec{}{j}$.
We solve Eq.\ (\ref{eq:equation of motion}) numerically
  using the Euler method with a fixed time interval of
  $5.48 \times 10^{-2} \sqrt{\mass{}{} / \springcoeff{}{}}$ for each step.

As mentioned above, we ignore the mutual deformation of the wall. Therefore, 
an embedded particle $i \in \setwps$ in the wall can be characterized by
its position in the radial direction $r_0(t;z)$ of $\posvec{}{i}
   = \bigl(
       \cylr{}{i} (\systime{}{}; \cylz{}{i}), \cylangle{}{i}, \cylz{}{i}
    \bigr)$
  in cylindrical coordinates as
\begin{equation}
  \cylr{}{i}(\systime{}{}; \cylz{}{i})
  = \biggl( \avgtuberadius + \dfrac{\diameter{}{}}{2} \biggr)
    + \perisamplitude
      \sin \dfrac{2\pi}{\wavelength}(\perisvel\systime{}{} + \cylz{}{i})
\label{eq:sinusoidal motion}
\end{equation}
  for a sinusoidal peristaltic wave of the wall
  with amplitude $\perisamplitude$, wavelength $\wavelength$,
  phase velocity $\perisvel$, and average tube radius $\avgtuberadius$.
In contrast, $\cylangle{}{i}$ and $\cylz{}{i}$ are set
  to form a nearly hexagonal lattice with lattice constant $l$
  on the $\cylangle{}{}$-$\cylz{}{}$ plane.
See the appendix for more precise descriptions of $\cylangle{}{i}$ and $\cylz{}{i}$.
We  fix $l = 0.3 \, \diameter{}{}$ throughout this \articletype{}
  to prevent the transported particles from penetrating the tube wall,
  which means that $\numwps = 9984$, although we set $\numps = 50, 100, \ldots, 600$ in our simulations.

Hereafter, we use dimensionless quantities scaled by
  the units of mass $\mass{}{}$, length $\diameter{}{}$,
  and time $\sqrt{\mass{}{} / \springcoeff{}{}}$.
For example, the length of the tube, time, and viscosity in dimensionless forms
  are given by $\tubelength / \diameter{}{}$,
  $\systime{}{} \sqrt{\springcoeff{}{} / \mass{}{}}$,
  and $\viscosity{}{} / \sqrt{\mass{}{} \springcoeff{}{}}$, respectively.
Note that we do not introduce any new notations
  for the dimensionless quantities in the following,
  and use the same notations as those for the dimensional quantities.

We set
  $\avgtuberadius = 1.5$, $\tubelength = 50$, $\wavelength = 10$,
  and $\viscosity{}{} = 5.48 \times 10^{-3}$ throughout this \articletype.
The controllable parameters are
  $\perisamplitude$, $\perisvel$, and $\numps$,
  but instead we use
  the minimum width of the peristaltic tube, \ie, the width at the bottleneck,
\begin{equation}
  \minwidth \equiv 2(\avgtuberadius - \perisamplitude),
\end{equation}
  the strain rate
\begin{equation}
  \strainrate \equiv \dfrac{\perisvel}{\wavelength},
\end{equation}
  and the volume fraction at $\perisamplitude = 0$,
\begin{equation}
  \effdens \equiv \dfrac{\numps}{6 \avgtuberadius{}^2 \tubelength},
\end{equation}
  as the control parameters in this work.
Note that we do not use the volume fraction
  $\dens \equiv
   \numps / 6 (\avgtuberadius{}^2 + \perisamplitude{}^2 / 2) \tubelength$
  but use $\effdens$ instead.
This is because it is difficult to fix $\dens$
  in simulations and in real experiments when $\perisamplitude$ is changed.
It is necessary to change both $\numps$ and $\perisamplitude$ at the same time. 
Moreover, $\perisamplitude$ must be changed discretely
  because $\numps$ is a natural number.
In this study, the restitution coefficient of the particles, $e$,
  is given by
  $e = \exp \bigl( -\pi \viscosity{}{} / \sqrt{2 - \viscosity{}{}^2} \bigr)
     \simeq 0.988$,
  \ie, the particles are almost elastic \cite{otsuki:2010}.
Note, however, that such ``low'' inelasticity or dissipation
  is necessary to reach a steady state because the energy is input to the system
  continuously by the peristaltic motion.

\section{Results}\label{sec:results}

We focus on the behavior of the flow rate $\massflux(\systime{}{})$,
whose explicit definition will be given in Sec.\ \ref{sec:definition},
  under the zero-flow-rate initial condition $\massflux(0) = 0$.
If $\minwidth$ is sufficiently large, the particles can flow without becoming stuck.
We call this an unjammed flow  (see Fig.\ \ref{fig:model} (b)).
On the other hand, if $\minwidth$ is sufficiently small, 
  particles become stuck at bottlenecks, and the stuck particles rise with the wall.
We call this a jammed flow (see Fig.\ \ref{fig:model} (c)).
We are interested in the particle behavior for intermediate values of $\minwidth$.
In Sec.\ \ref{sec:transition}, we present the results
  obtained by fixing $\effdens$ and changing $\minwidth$ and $\strainrate$.
Meanwhile, the results obtained by changing $\effdens$
  are given in Sec.\ \ref{sec:density}.
Finally, the results for the second moment of the flow rate are reported
  in Sec.\ \ref{sec:susceptibility}.

It is noteworthy that the terminology of ``jammed" and ``unjammed" flows is counter-intuitive, because
a jammed flow has a larger flow rate than an unjammed flow.
This can be understood easily by considering the process relative to a frame moving with the wall.
Indeed, the jammed state does not exhibit any motion in this frame, 
while the unjammed state has some mobile particles moving in the backward direction. 
Namely, some particles left behind by the wall motion contribute to the negative flow rate 
in the unjammed phase.

\subsection{Definition of flow rate}\label{sec:definition}

\begin{figure*}
  \begin{center}
    \includegraphics[width=.45\linewidth]{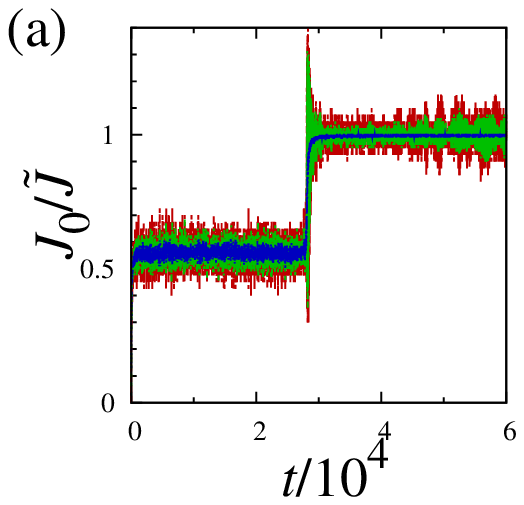}
    \includegraphics[width=.45\linewidth]{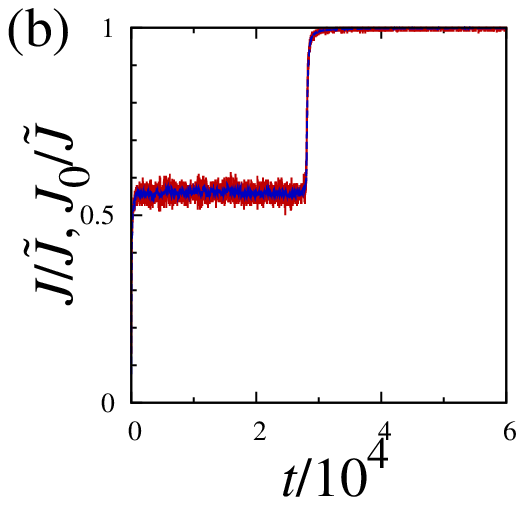}
    \\
    \includegraphics[width=.45\linewidth]{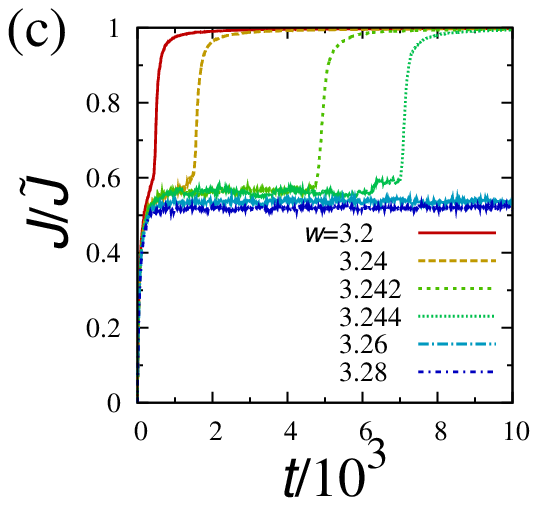}
    \includegraphics[width=.45\linewidth]{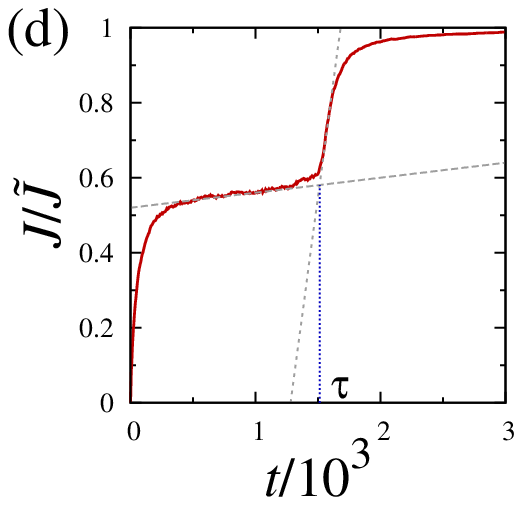}
  \end{center}
  \caption{\label{fig:time evolution}
  (Color online)
  (a) Time evolutions of the normalized
    flow rate $\massflux_0(\systime{}{}, \Delta \systime{}{}) / \maxmassflux$
     with $\tilde{J}\equiv N c/L$    
for $\minwidth = 2.2448$, $\strainrate = 1.83 \times 10^{-1}$,
    and $\effdens = 2.96 \times 10^{-1}$.
  The definition of $\massflux_0(\systime{}{}, \Delta \systime{}{})$ is given
    by Eq.\ (\ref{eq:another flow rate}).
  The observation times $\Delta \systime{}{}$
    for $\massflux_0(\systime{}{}, \Delta \systime{}{})$ are set
    to $T$, $2T$, and $5T$,
    where $T = 5.48$ is the period of the peristaltic wave.
  (b) Time evolutions of the normalized
    flow rates $\massflux(\systime{}{}) / \maxmassflux$
    and $\massflux_0(\systime{}{}, 5T) / \maxmassflux$.
    The parameters are the same as those in (a).
  (c) Typical time evolution of normalized flow rate $\massflux / \maxmassflux$
    with $\strainrate = 1.83 \times 10^{-1}$
    and $\effdens = 2.96 \times 10^{-1}$.
  (d) Same plot as (c) for $\minwidth = 3.24$ to show the definition of the transition time $\transtime$.
  See text for details.}
\end{figure*}

The flow rate is often defined \cite{evans:1990} as
\begin{equation}
  \massflux(\systime{}{})
  \equiv \dfrac{1}{\tubelength}
         \sum_{i \in \setps} \mom{}{i, \cylz{}{}}(\systime{}{})
\label{eq:flow rate}
\end{equation}
  in studies employing molecular dynamics simulations and discrete element methods,
  where $\mom{}{i, \cylz{}{}}(\systime{}{})$ is
  the $\cylz{}{}$-component of the dimensionless momentum of particle $i$.
In real experiments, however, the above definition may not be appropriate
  because it requires the momenta of all particles.
Instead, the following definition is usually used:
\begin{equation}
  \massflux_{\cylz{}{}'}(\systime{}{}, \Delta \systime{}{})
  \equiv \dfrac{\Delta \numps_{\cylz{}{}'}(\systime{}{}, \Delta \systime{}{})}
               {\Delta \systime{}{}},
\label{eq:another flow rate}
\end{equation}
  where $\Delta \numps_{\cylz{}{}'}(\systime{}{}, \Delta \systime{}{})$ is
  a signed number of particles passing a section $\cylz{}{} = \cylz{}{}'$
  during the time interval $[\systime{}{}, \systime{}{} + \Delta \systime{}{})$,
  with $+1$ counted if a particle passes from $\cylz{}{} < \cylz{}{}'$
  to $\cylz{}{} > \cylz{}{}'$ and $-1$ counted for the opposite case.
Therefore, we first check that the definition given by Eq.\ (\ref{eq:flow rate}) is
  in agreement with that given by Eq.\ (\ref{eq:another flow rate}).

Figure \ref{fig:time evolution} (a) plots
  time evolutions of the flow rate
  $\massflux_{z'=0}(\systime{}{}, \Delta \systime{}{})$
  normalized by $\maxmassflux = \numps \perisvel / \tubelength$. 
Note that if all particles rise with the wall, \ie, $\velvec{}{i} = (0, 0, \perisvel)$ for every particle $i$,
$\massflux$ is equal to $\maxmassflux$.  
It is clearly shown that
  the fluctuation of the curve decreases as $\Delta \systime{}{}$ increases.
Moreover, Fig.\ \ref{fig:time evolution} (b) shows
  time evolutions of the normalized flow rates
  $\massflux(\systime{}{}) / \maxmassflux$
  and $\massflux_0(\systime{}{}, 5T) / \maxmassflux$,
  where $T \equiv \wavelength / \perisvel = 1 / \strainrate$.
From these figures, we confirm  that 
   $\massflux_0(\systime{}{}, \Delta t)$ converges to $\massflux(\systime{}{})$ for sufficiently large $\Delta t$.
Therefore, if one is not interested in short-time behavior,
  Eq.\ (\ref{eq:flow rate}) can be used for data analysis.
Hereafter, we adopt Eq.\ (\ref{eq:flow rate}) for the flow rate in the subsequent discussion.

\subsection{Transition between jammed and unjammed flow}\label{sec:transition}

Figure \ref{fig:time evolution} (c) shows a typical example of the time evolution
  of the flow rate $\massflux / \maxmassflux$ for various $\minwidth$.
It is clearly shown that,
  while an almost stationary and slow unjammed flow continues for large $\minwidth$,
  a transition from a slow unjammed flow to a fast jammed flow occurs
  for small $\minwidth.$

We introduce the transition time $\transtime$
  to characterize the transition from the unjammed flow to the jammed flow
  (Fig.\ \ref{fig:time evolution} (c)).
First, we linearly fit the curve of the flow rate
  before a transition (gentle dashed line).
We again fit the curve after the transition in a similar way (steep dashed line).
Then the transition time is defined as the time
  at which these two fitted lines cross.
Note that the transition time in this definition is
  not the lifetime of a metastable state
  but the relaxation time  from the start of the unjammed flow.
Such a definition is commonly used
  to characterize the relaxation from metastable states
  in melting dynamics \cite{binder:1973}
  and the so-called $\alpha$-relaxation
  in glass transitions \cite{kob:1994,kob:1995}.

\begin{figure}
  \begin{center}
    \includegraphics[width=.45\linewidth]{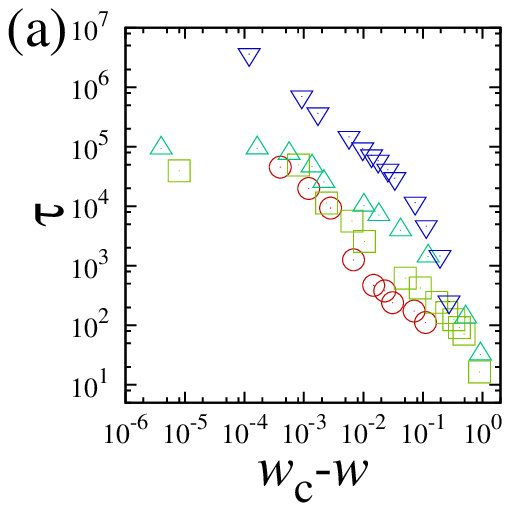}
    \includegraphics[width=.45\linewidth]{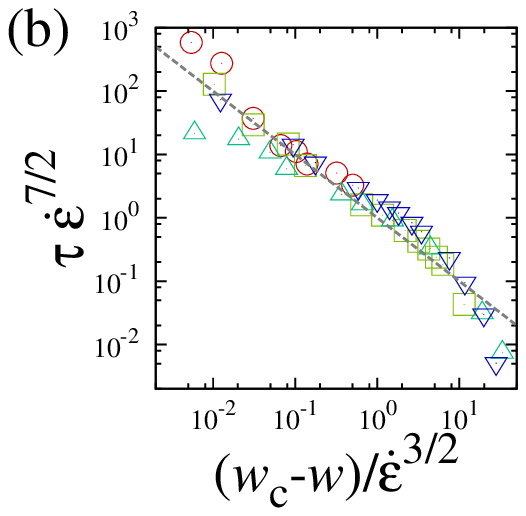}
    \\
    \includegraphics[width=.45\linewidth]{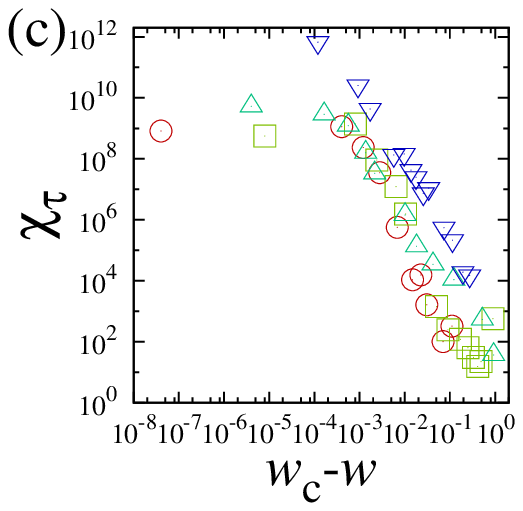}
    \includegraphics[width=.45\linewidth]{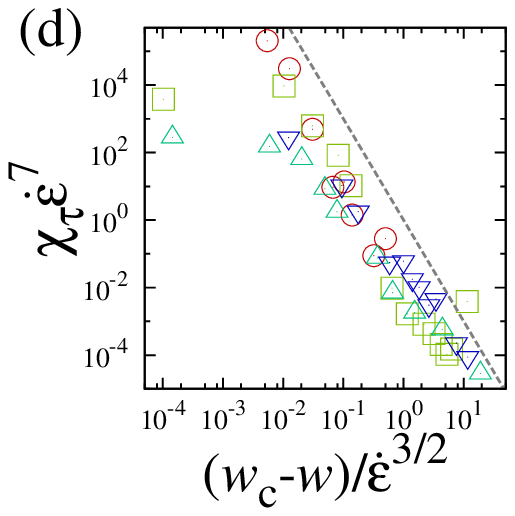}
  \end{center}
  \caption{\label{fig:time}
  (Color online)
  (a) Transition time $\transtime$
    as a function of $\critminwidth - \minwidth$.
  (b) Scaling function $f(x)$
    with $x = (\critminwidth - \minwidth) / \strainrate{}^{\nu}$
    and $\nu = 3/2$;
  see Eq.\ (\ref{eq:t scaling}).
  (c) Fluctuation of transition time $\transtimefluc$
    as a function of $\critminwidth - \minwidth$.
  (d) Scaling function $g(x)$
    with $x = (\critminwidth - \minwidth) / \strainrate{}^{\nu}$;
  see Eq.\ (\ref{eq:chi_t scaling}).
  For all figures, we use the density $\effdens = 2.96 \times 10^{-1}$,
    and circles, squares, triangles, and inverted triangles correspond to the data for
    $\strainrate = 3.65 \times 10^{-1}, 1.83 \times 10^{-1},
     9.13 \times 10^{-2}$, and $4.56 \times 10^{-2}$, respectively.}
\end{figure}

Figure \ref{fig:time} (a) shows
  the $\minwidth$-dependence of $\transtime$ for various $\strainrate$, which suggests the existence of a power law
between $\tau$ and $w_c-w$.
  Although it may be difficult to extract a universal feature from Fig. \ref{fig:time} (a), 
the data may satisfy the scaling form
\begin{equation}
  \transtime
  \simeq \strainrate{}^{-\zeta}
       f\bigl( (\critminwidth - \minwidth) / \strainrate{}^{\nu} \bigr),
\label{eq:t scaling}
\end{equation}
  where $f(x) \sim x^{-\transtimeexp}$ for $x \sim 1$
  and might undergo exponential decay for $x \gg 1$ (see Fig.\ \ref{fig:time} (b)).
Note that the critical width $w_c$ is determined from Eq. (\ref{eq:t scaling}).
This is because the raw data obtained from the simulation, as shown in Fig.\ \ref{fig:time} (a), cannot reach the critical point within our simulation time. 
From Fig. \ref{fig:time} (b), we numerically obtain $\transtimeexp = 1$, $\nu = 3/2$,
  and $\zeta = 7/2$ for $\effdens = 2.96 \times 10^{-1}$.
The existence of the scaling law given by Eq. (\ref{eq:t scaling}) suggests 
  that this jamming transition is analogous to the conventional second-order phase transition.

We also study the $\minwidth$-dependence of the fluctuation of the transition time
  $\transtimefluc
   \equiv \average{\transtime{}^2} - \average{\transtime}^2$.
We again find, as shown in Figs.\ \ref{fig:time} (c) and (d),
  the existence of the scaling form
\begin{equation}
  \transtimefluc
  \simeq \strainrate{}^{-2\zeta}
       g\bigl( (\critminwidth - \minwidth) / \strainrate{}^{\nu} \bigr),
\label{eq:chi_t scaling}
\end{equation}
  where $g(x) \sim x^{-\beta}$ for $x \sim 1$
  and might undergo exponential decay for $x \gg 1$.
The value of $\beta$ is approximately equal to $3$ for $\effdens = 2.96 \times 10^{-1}$.

\begin{figure*}
  \begin{center}
    \includegraphics[width=.45\linewidth]{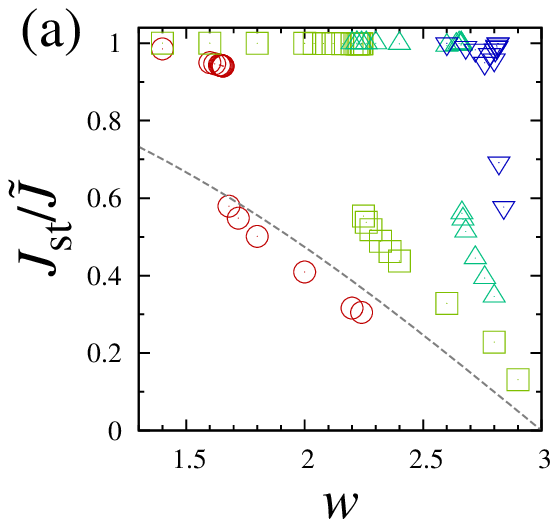}
    \\
    \includegraphics[width=.45\linewidth]{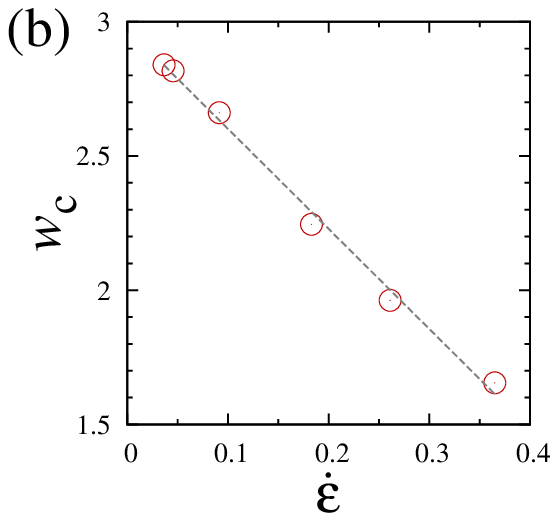}
    \includegraphics[width=.45\linewidth]{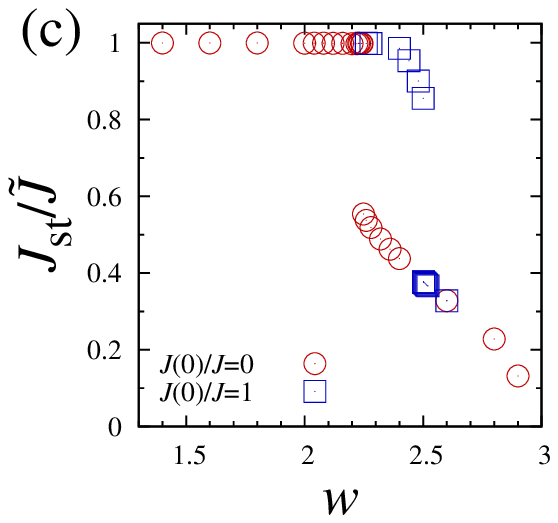}
  \end{center}
  \caption{\label{fig:flux 1}
  (Color online)
  (a) Normalized stationary flow rate $\stmassflux / \maxmassflux$
    as a function of $\minwidth$ for $\effdens = 2.96 \times 10^{-1}$
    and various $\strainrate$,
    where the dashed line is given by Eq.\ (\ref{eq:flow rate stokes}).
  The symbols are the same  as those in Fig.\ \ref{fig:time}.
  (b) Plot of $\critminwidth$ as a function of $\strainrate$
    for $\effdens = 2.96 \times 10^{-1}$.
  (c) Hysteresis loop of flow rate for $\effdens = 2.96 \times 10^{-1}$ and $\strainrate = 1.83 \times 10^{-1}$.
  Circles and squares correspond to
    $\stmassflux / \maxmassflux = 0$ and $1$, respectively.}
\end{figure*}

We now discuss how $\critminwidth$ depends on $\strainrate$.
Figure \ref{fig:flux 1} (a) shows
  the $\minwidth$-dependence of $\stmassflux / \maxmassflux$,
  where $\stmassflux$ is
the stationary flow rate of the jammed flow for $\minwidth < \critminwidth$ or that of the unjammed flow
   for $\minwidth > \critminwidth$.
There is a jump in each stationary flow rate $\stmassflux$
  at $\minwidth = \critminwidth(\strainrate)$.
The transition point $\critminwidth(\strainrate)$
  strongly depends on $\strainrate$ as shown in Fig.\ \ref{fig:flux 1} (b).
This figure shows that
  $\critminwidth$ linearly increases with decreasing $\strainrate$ as
\begin{equation}
  \critminwidth(\strainrate)
  \simeq -3.75 \strainrate + \upperminwidth,
\end{equation}
  at least for $\strainrate<0.4$.

The jump in the stationary flow rate may imply
that  the jamming transition is not continuous
  but discontinuous, although the scaling forms given by Eqs.\ (\ref{eq:t scaling}) and (\ref{eq:chi_t scaling}) are usually a characteristic of a continuous transition.
To determine whether our jamming transition is continuous or discontinuous, we have investigated whether there is  hysteresis in the vicinity of the transition point. 
Figure \ref{fig:flux 1} (c) illustrates the existence of a hysteresis loop of the flow rate
  for $\effdens = 2.96 \times 10^{-1}$ and $\strainrate = 1.83 \times 10^{-1}$,
  which is clear evidence that this jamming transition is discontinuous.
In the figure, squares and circles correspond to data obtained from simulations
  with the initial conditions $\massflux(0) = \maxmassflux$ and $J(0)=0$, respectively.
For $J(0)=\tilde{J}$, 
  there is a stable and steady jammed flow for $\minwidth < \newcritminwidth$,
  a transition from the jammed flow to the unjammed flow takes place at $w=w_c'$,
  and the unjammed flow is stable for $\minwidth > \newcritminwidth$.
It is clear that the transition point  $\newcritminwidth$ satisfies  $w_c'> \critminwidth$.

It is well known that
  the transition time or relaxation time
  from a metastable state to a stable equilibrium state 
  diverges at a critical point
  for a discontinuous phase transition
  \cite{binder:1973,binder:1987,krzakala:2011a,krzakala:2011b}.
One commonly observes that the Vogel-Fulcher-Tammann law,
  or its variant $\tau \sim \exp (-C\delta^{-\gamma})$, is satisfied for the relaxation time,
  where $\delta$ is a rescaled control parameter or the distance from the critical or spinodal point. 
Meanwhile, 
  the power-law divergence
  $\tau \sim \delta^{-\alpha}$ is also often observed
  near the spinodal point in mean-field models
  \cite{binder:1973,binder:1987,krzakala:2011a,krzakala:2011b}
  and in molecular dynamics simulations for 
 sheared colloidal systems
  \cite{miyama:2011}.
Our system is another example of a system
  in which
  the transition time to a stable state exhibits
  power-law divergence at a critical point as $\tau \sim \delta^{-\alpha}$, with $\delta = 1 - \minwidth/\critminwidth$ and $\alpha=1$
  for a nonequilibrium discontinuous phase transition.

We also investigate the dependence of the stationary flow rate on $\minwidth$ in the jammed and unjammed flow phases.
The dashed curve in Fig.\ \ref{fig:flux 1} (a) shows the analytic relation between the stationary flow rate $J_{\rm st}^{\rm fluid}$
  and the minimum width in a Stokes fluid under an infinite wavelength limit
  \cite{jaffrin:1971},
\begin{equation}
  \dfrac{\fluidstmassflux}{\maxmassflux}
  = \dfrac{3\phi}{2 + \phi^2}
  = \dfrac{3 (1 - \minwidth /\avgtuberadius)}
          {2 + (1 - \minwidth /\avgtuberadius)^2},
\label{eq:flow rate stokes}
\end{equation}
  where 
  $\phi = \perisamplitude / \avgtuberadius = 1 - \minwidth / \avgtuberadius$.
In the jammed flow phase with  $\minwidth < \critminwidth$,
  the stationary flow rate is almost independent of $\minwidth$
  and, as expected, does not follow Eq.\ (\ref{eq:flow rate stokes}).
On the other hand, the flow rate decreases as $w$ increases 
 in the unjammed flow phase with $\minwidth > \critminwidth$.
  It is surprising that 
  the functional form of the stationary flow rate is similar to Eq.\ (\ref{eq:flow rate stokes})
   for large $\dot\epsilon$, \eg, $\dot\epsilon=0.365$ (see Fig. \ref{fig:flux 1} (a)), 
although the flow rate is larger for smaller $\dot\epsilon$.
This is counter-intuitive because the Reynolds number is usually defined as $Re = \avgtuberadius^2 \perisvel / \nu \wavelength$ with kinematic viscosity $\nu$, and thus,
  the flow rate converges to Eq.\ (\ref{eq:flow rate stokes}) as $\strainrate \to 0$ in the case of a Stokes flow \cite{jaffrin:1971, shapiro:1969, weinberg:1971, jaffrin:1973}.
A means of solving this puzzle might be to consider the effect of compressibility.
Indeed, it is known that the stationary flow rate increases as $\perisvel$ decreases
  if $\perisvel$ is greater than the sound velocity \cite{aarts:1998,felderhof:2011}.
It is also known that a granular fluid is compressible \cite{brilliantov:2004}.
Of course, the existence of a role of incompressibility raises an interesting question:
  why is the behavior of a suspension flow in an incompressible fluid similar to that of a Stokes flow?
At present, this is an open question.

\subsection{Density dependence}\label{sec:density}

\begin{figure}
  \begin{center}
    \includegraphics[width=.45\linewidth]{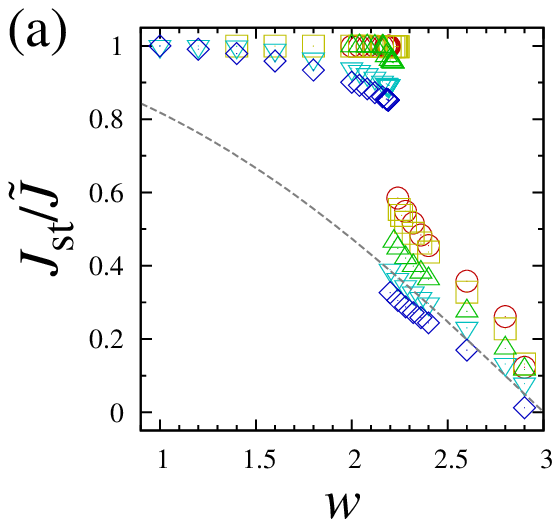}
    \includegraphics[width=.45\linewidth]{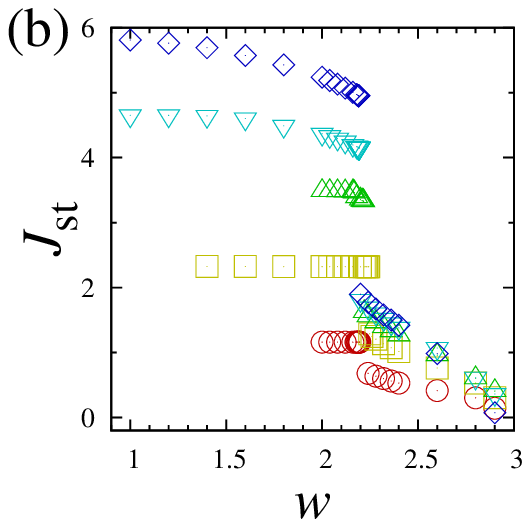}\\
    \includegraphics[width=.45\linewidth]{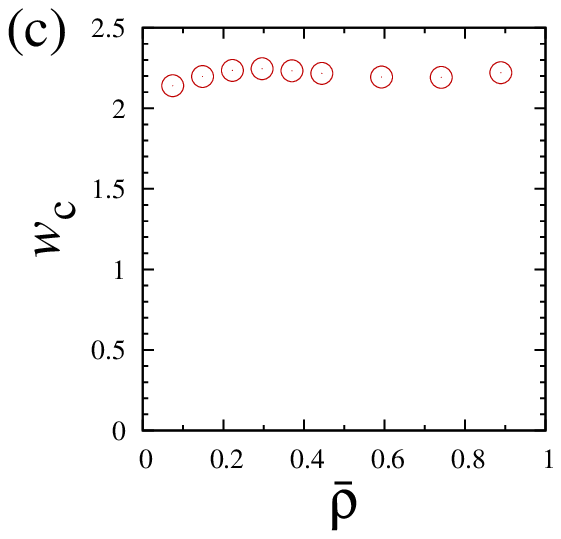}
    \includegraphics[width=.45\linewidth]{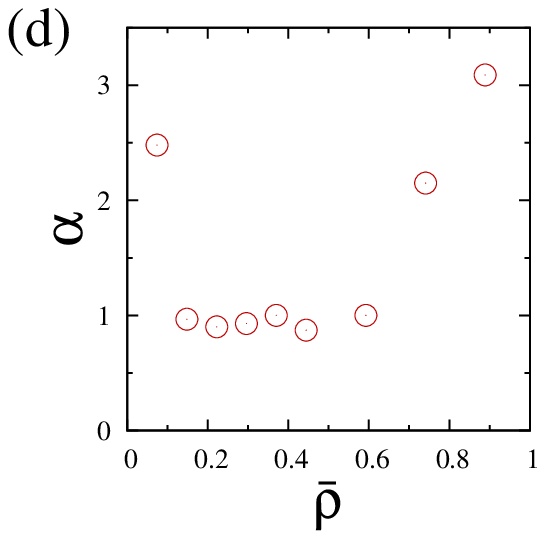}
  \end{center}
  \caption{\label{fig:flux 2}
  (Color online)
  (a) Normalized stationary flow rate $\stmassflux / \maxmassflux$
    as a function of $\minwidth$ for various $\effdens$.
    The dashed line shows Eq.\ (\ref{eq:flow rate stokes}).
  Circles, squares, triangles, inverted triangles, and diamonds
    correspond to
    $\effdens = 1.48 \times 10^{-1}, 2.96 \times 10^{-1},
     4.44 \times 10^{-1}, 5.92 \times 10^{-1}$, and $7.40 \times 10^{-1}$,
    respectively.
  (b) Unnormalized stationary flow rate.
  (c) Plot of $\critminwidth$ as a function of $\effdens$.
  (d) Plot of $\transtimeexp$ as a function of $\effdens$.
  All figures are obtained for $\strainrate = 1.83 \times 10^{-1}$.}
\end{figure}

In this subsection
  we investigate how the flow rate depends on the density under a fixed strain rate
  of $\strainrate = 1.83 \times 10^{-1}$.
Figures \ref{fig:flux 2} (a) and (b) show the normalized and unnormalized stationary flow rates $\stmassflux / \maxmassflux$ and 
$\stmassflux$ as functions of $\minwidth$ for various $\effdens$, respectively.
Although the actual flow rate $J_{\rm st}$ increases as the density increases (Fig. \ref{fig:flux 2} (b)),
  the normalized flow rate decreases with increasing density (Fig. \ref{fig:flux 2} (a)).
For the jammed flow phase, $\minwidth < \critminwidth$, with relatively large $\effdens$,
  there are some particles that do not move with the wall.
Note that the transition point $\critminwidth\simeq 2.2$ is almost independent of the density
  as shown in Fig.\ \ref{fig:flux 2} (c).

\begin{figure}
  \begin{center}
    \includegraphics[height=.6\linewidth]{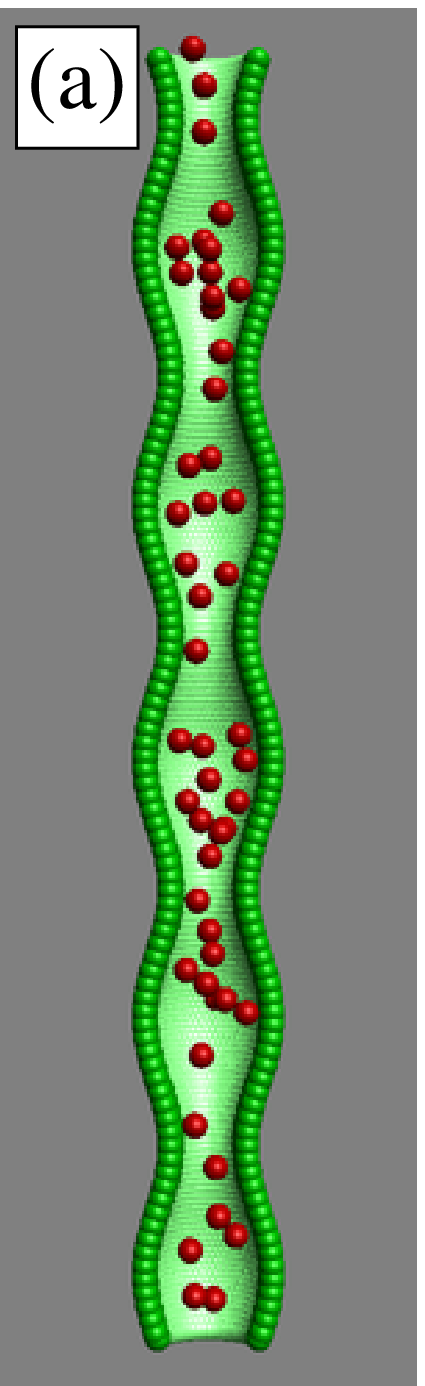}
    \includegraphics[height=.6\linewidth]{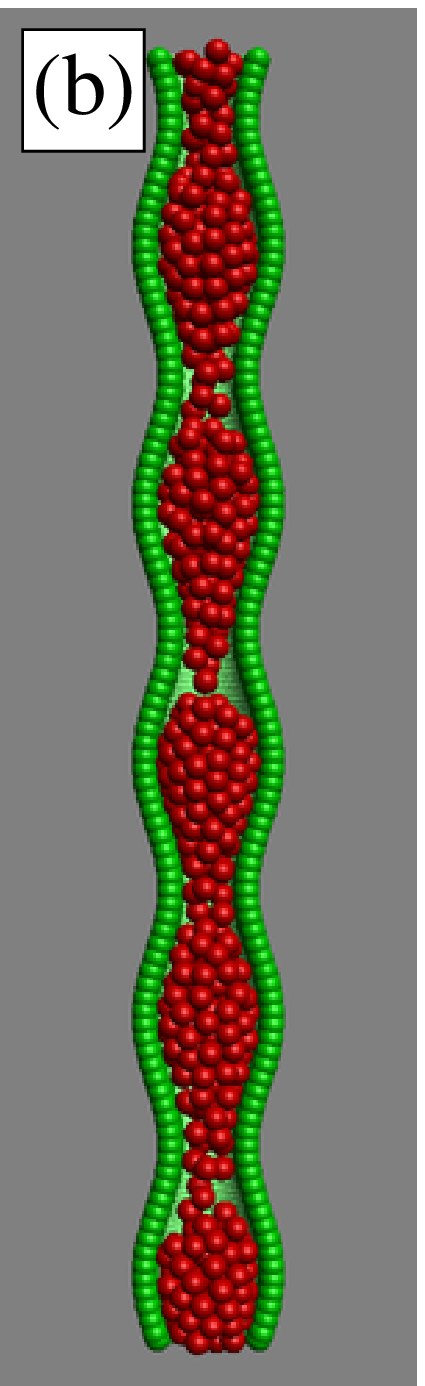}
  \end{center}
  \caption{\label{fig:density dep}
  (Color online)
  (a) Snapshot of stationary state
    for $\minwidth = 2.0$, $\strainrate = 1.83 \times 10^{-1}$,
    and $\effdens = 7.40 \times 10^{-2}$.
  Although particles do not become stuck at bottlenecks,
  the stationary flow rate $\stmassflux$ reaches $\maxmassflux$.
  (b) Snapshot of stationary state
    for $\minwidth = 2.0$, $\strainrate = 1.83 \times 10^{-1}$,
    and $\effdens = 7.40 \times 10^{-1}$.
}
\end{figure}

We also demonstrate that
  the exponent $\transtimeexp$ is a constant for $0.15 \lesssim \effdens \lesssim 0.60$ in Fig.\ \ref{fig:flux 2} (d).
For a dilute gas with $\effdens \lesssim 0.10$, $\transtimeexp$ is much larger than $1$.
This might be because the transport of particles is not determined by 
  the rare collisions between particles
  but by the direct momentum transfer from the peristaltic wave of the wall
  (Fig.\ \ref{fig:density dep} (a)).
On the other hand,
 $\transtimeexp$ is also larger than unity for $\effdens \gtrsim 0.70$.
This might be related to the insufficient free volume around the antinodes of the tube
  (Fig.\ \ref{fig:density dep} (b)).
Particles passing through a bottleneck soon collide with other jammed particles
  around the next antinode and thus they are deflected.
These deflected particles may affect
  the configuration of jammed particles near the bottleneck.
Because  we focus on the appearance of the jammed flow,
  we can state that the relationship $\transtimeexp \simeq 1$ is always held for a wide range of densities.

The dashed curve in Fig.\ \ref{fig:flux 2} (a) shows Eq.\ (\ref{eq:flow rate stokes})
	or the $\minwidth$-dependence of the flow rate in a Stokes fluid.
We find that
	the flow rate in the unjammed flow phase becomes increasingly close
	to that given by Eq.\ (\ref{eq:flow rate stokes}) with increasing density $\effdens$.
This also indicates that compressibility has a crucial role
	in causing the functional form of $\stmassflux / \maxmassflux$
	to deviate from Eq.\ (\ref{eq:flow rate stokes}).
It is also worth noting that
	$\stmassflux / \maxmassflux$ is not constant in the jammed phase
	if $\effdens$ is sufficiently high.

\subsection{Second moment of flow rate}\label{sec:susceptibility}

\begin{figure}
  \begin{center}
    \includegraphics[width=.45\linewidth]{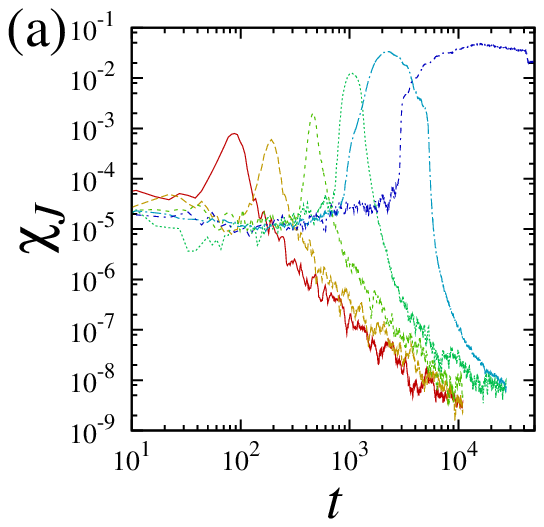}
    \includegraphics[width=.48\linewidth]{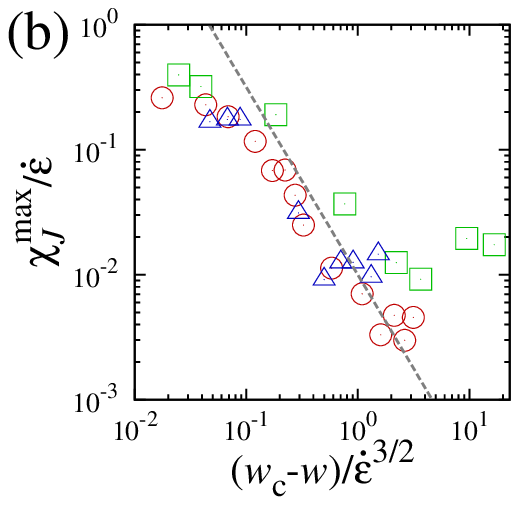}
  \end{center}
  \caption{\label{fig:dynamic susceptibility}
  (Color online)
  (a) Second moment of the flow rate $\dynsuscept{}{\massflux}$
    as a function of $\systime{}{}$.
  We set $\strainrate = 1.8 \times 10^{-1}$,
    and curves from left to right correspond to
    $\minwidth = 2.0, 2.12, 2.2, 2.232, 2.24$, and $2.244$.
  (b) Scaling function $h(x)$ of peak values of $\dynsuscept{}{\massflux}$,
    $\maxdynsuscept{\massflux}$,
    with $x = (\critminwidth - \minwidth) / \strainrate{}^{\nu}$;
  see Eq.\ (\ref{eq:chi^max scaling}).
  Circles, squares, and triangles correspond to
    $\strainrate = 1.83 \times 10^{-1}, 9.13 \times 10^{-2}$,
    and $4.56 \times 10^{-2}$, respectively.
  Both figures are obtained for $\effdens = 2.96 \times 10^{-1}$.}
\end{figure}

Finally, we measure the second moment of the normalized flow rate,
  $\dynsuscept{}{\massflux} (\systime{}{})
   \equiv (\average{\massflux(\systime{}{})^2}
          - \average{\massflux(\systime{}{})}^2) / \maxmassflux{}^2$.
From this quantity, we might detect growing length and time scales near the dynamical phase transition.
Indeed, Krzakala and Zdeborov\'a have recently analyzed dynamic susceptibility in order to characterize
  the melting dynamics leading to the \textit{equilibrium} state of spin-glasses
  \cite{krzakala:2011a,krzakala:2011b}.
They demonstrated  that there is a growing length scale near
  the critical point
  even in a discontinuous phase transition. 

Figure \ref{fig:dynamic susceptibility} (a) plots
  the time evolution of $\dynsuscept{}{\massflux} (\systime{}{})$
  for various $\minwidth$.
Each $\dynsuscept{}{\massflux}$ exhibits a peak at an intermediate time
  then vanishes with increasing time.
The peak value $\maxdynsuscept{\massflux}$ increases
  as $\minwidth$ approaches $\critminwidth$.
As shown in Fig.\ \ref{fig:dynamic susceptibility} (b),
  the data for $\maxdynsuscept{\massflux}$ satisfy the  scaling form
\begin{equation}
  \maxdynsuscept{\massflux}
  \sim \strainrate{}^{\xi}
       h\bigl( (\critminwidth - \minwidth) / \strainrate{}^{\nu} \bigr),
\label{eq:chi^max scaling}
\end{equation}
  where $h(x) \sim x^{-\gamma}$ for $x \sim 1$
  and $\xi = 1$, $\gamma = 3/2$.

\section{Concluding remarks}\label{sec:summary}

We demonstrated that
  the flow rate is a suitable order parameter for this system
  and found that a dynamic phase transition from a slow ``unjammed" flow to a fast ``jammed" flow
  occurs at a critical width of the peristaltic tube.
We also found that the transition times, its fluctuations,
  and the peak values of the second moments of the flow rate
  obey power laws
  as the minimum width approaches the critical value.
Moreover, we demonstrated the existence of scaling functions for these quantities.
It was also shown that
  the critical width is almost independent of the density
  but depends linearly on peristaltic velocity.
Nevertheless, the phase transition is discontinuous and exhibits a hysteresis loop under some initial conditions.

To the best of the authors' knowledge, this is the first report
  demonstrating the existence of a phase transition in peristaltic transport.
Since the pioneering work on peristaltic transport 
  carried out by Shapiro \etal\ \cite{shapiro:1969},
  many works have reported how the flow rate depends on the amplitude ratio
  $\phi \equiv \perisamplitude / \avgtuberadius
   = 1 - \minwidth / 2\avgtuberadius$ for various fluids.
In contrast, our work reveals that
  the flow rate of a granular flow exhibits a finite jump
  at $\phi = \phi_\mathrm{c} = 1 - \critminwidth / 2 \avgtuberadius$.
To describe this behavior using a fluid model, the authors suggest that
  at least compressibility and the local fluid-solid phase transition have to be introduced into the model.
We expect that our finding will be confirmed by experiments in the near future.

As mentioned in Sec.\ \ref{sec:introduction},
  an analogous phase transition exists
  in a granular flow through a small bottleneck,
  which is called the dilute-to-dense transition
  \cite{hou:2003, zhong:2006, huang:2006}.
Recently Zhong \etal\ reported \cite{zhong:2006} that
  the characteristic time $\tau$ satisfies $\tau\sim (w_c-w)^{-\alpha}$ with $\alpha=1.8 \pm 0.2$
  in a two-dimensional granular flow through a bottleneck.
Their exponent is, of course, much larger than
  ours of $\transtimeexp = 1$.
There is no reason why the same exponent should be obtained in different systems with different dimensions, 
but we expect that there is a universal law that explains such dynamical phase transitions. 

In this \articletype, we ignored many aspects of realistic granular particles such as
  the static friction among contact grains, the polydispersity of grain size, 
  particle shape, and the effects of adhesive forces and interstitial fluids.
Moreover, we stress that the strain-control protocol used in this \articletype{} is unrealistic
  and that a stress-control protocol should be used for a realistic analysis of the peristaltic transport of grains.
In this sense, our paper is only the first step to demonstrating the possible existence of an interesting
  dynamic phase transition for the peristaltic transport of granular particles,
  which might have occurred as a result of the oversimplified model that we used. 

\begin{acknowledgments}
This work was partially supported
  by the Ministry of Education, Culture, Sports, Science and Technology of Japan (MEXT)
  (Grant No.\ 21540384)
  and by a Grant-in-Aid for the Global COE Program
  ``The Next Generation of Physics, Spun from Universality and Emergence''
  from MEXT.
N.Y. was supported by a Grant-in-Aid for JSPS Fellows.
The numerical calculations were partly carried out on the Altix3700 BX2 supercomputer
  at Yukawa Institute for Theoretical Physics at Kyoto University.
\end{acknowledgments}

\appendix*

\section{Configuration of particles embedded in a wall}\label{sec:appendix}

It was explained in Sec.\ \ref{sec:model}
  that the peristaltic tube is constructed from particles
  with identical material properties
  and that their radial motion $\cylr{}{i}(\systime{}{}; \cylz{}{i})$
  is given by Eq.\ (\ref{eq:sinusoidal motion}).
In this appendix, we determine the azimuth and height components
  $(\cylangle{}{i}, \cylz{}{i})$ of the position of each particle
  $i \in \setwps = \{ \numps + 1, \ldots, \numps + \numwps \}$.

\begin{figure}
  \begin{center}
    \includegraphics[width=.45\linewidth]{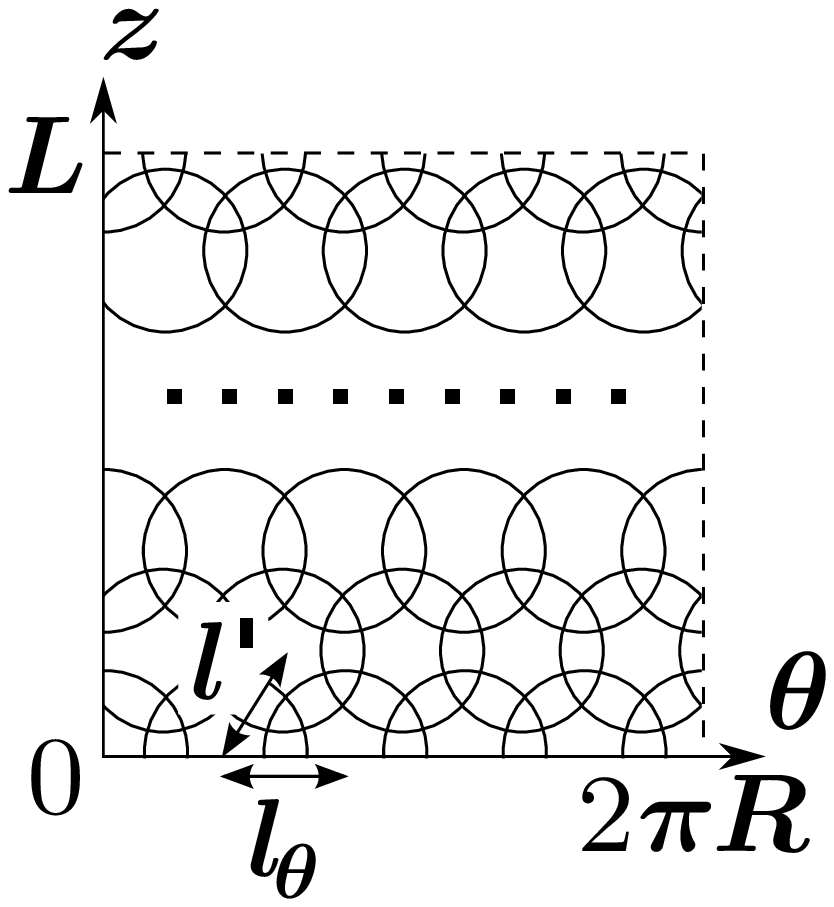}
  \end{center}
  \caption{\label{fig:phyz}
  Schematic diagram of the configuration of wall particles in the $\cylangle{}{}$-$\cylz{}{}$ plane.
  Here, $R = \avgtuberadius + \diameter{}{} / 2$.}
\end{figure}

In Sec.\ \ref{sec:model},
  we described the configuration of the wall particles as being nearly hexagonal.
We first explain what ``nearly hexagonal'' means.
Figure \ref{fig:phyz} shows
  a schematic diagram of the positions of wall particles
  in the $\cylangle{}{}$-$\cylz{}{}$ plane.
Here the particles form a nearly hexagonal lattice
  with lattice constant $l_\cylangle{}{}$ in the $\cylangle{}{}$-direction
  and lattice constant $l'$ in oblique directions.
The configuration is perfectly hexagonal if $l_\cylangle{}{} = l'$.
In our simulation setup, however, we set $l_\cylangle{}{} \neq l'$,
  and we consider this configuration to be nearly hexagonal.
This is because of the periodicity in the $\cylangle{}{}$- and $\cylz{}{}$-directions;
  if the configuration was perfectly hexagonal,
  the tube length $L$ and average tube radius $a$ would satisfy
  the relation $2\pi (\avgtuberadius + \diameter{}{} / 2) / n = \tubelength / \sqrt{3} \, n'$
  for $n, n' \in \mathbb{N}$.
To avoid imposing this constraint,
  we consider the above nearly hexagonal lattice in this \articletype.

We introduce a parameter $l$ to determine the lattice constants $l_\cylangle{}{} \simeq l$ and $l' \simeq l$.
We denote the number of particles in a layer (particles with identical $\cylz{}{}$ values) as $N'_\mathrm{w}$ and define the number of layers $N_\mathrm{l} = \numwps / N'_\mathrm{w}$ as
\begin{align}
  N'_\mathrm{w}
&
  = \biggl\lfloor
      \dfrac{2\pi (\avgtuberadius + \diameter{}{} / 2)}{l}
    \biggr\rfloor,
\label{eq:nw}
\\
  N_\mathrm{l}
&
  = 2\Biggl\lfloor
      \dfrac{L}{\sqrt{3} \, l}
    \Biggr\rfloor,
\label{eq:nl}
\end{align}
  where $\lfloor x \rfloor \equiv \max \{ n \in \naturalset | n \le x \}$
  is the integer part of $x$.
From Eqs.\ (\ref{eq:nw}) and (\ref{eq:nl}), $l_\cylangle{}{}$ and $l'$ are given as
\begin{align}
  l_\cylangle{}{}
  &= \dfrac{2\pi (\avgtuberadius + \diameter{}{} / 2)}{N'_\mathrm{w}}
\\
  l'
  &= \dfrac{2\sqrt{3}}{3} \dfrac{\tubelength}{N_\mathrm{l}}.
\end{align}


%

\end{document}